\documentclass[journal]{IEEEtran}

\usepackage{amssymb,amsmath,amsfonts,amsthm}
\usepackage{mathrsfs}
\usepackage{cite}
\usepackage{graphicx}
\usepackage{mathtools}
\usepackage{multirow}
\usepackage{color}
\usepackage{xcolor}
\usepackage{subfigure}
\DeclareMathOperator{\sinc}{sinc}

\IEEEoverridecommandlockouts

\IEEEoverridecommandlockouts
 
 \ifCLASSINFOpdf
 \else
 \fi

\begin{document}

\title{Probabilistic Caching in Wireless D2D Networks: Cache Hit Optimal vs. Throughput Optimal}

\author{Zheng~Chen, Nikolaos~Pappas and Marios~Kountouris
	\thanks{Z. Chen is with the Laboratoire de Signaux et Syst\`{e}mes (L2S, UMR8506),
		CentraleSup\'{e}lec - CNRS - Universit\'{e} Paris-Sud,
		Gif-sur-Yvette, France. Email: zheng.chen@centralesupelec.fr.}% <-this % stops a space
	\thanks{N. Pappas is with Department of Science and Technology, Link\"{o}ping University, Norrk\"{o}ping, Sweden. 
	Email: nikolaos.pappas@liu.se. }
	\thanks{M. Kountouris is with the Mathematical and Algorithmic Sciences Lab, France Research Center, Huawei Technologies Co., Ltd. Email: marios.kountouris@huawei.com. }
	\thanks{The research leading to these results has been partially funded by the European Union's Horizon 2020 research and innovation programme 
	under the Marie Sklodowska-Curie Grant Agreement No. 642743 (WiVi-2020).}% <-this % stops a space
	\vspace{-0.7cm}
}
\maketitle

\begin{abstract}
Departing from the conventional cache hit optimization in cache-enabled wireless networks, we consider an alternative optimization approach for the probabilistic caching placement in stochastic wireless D2D caching networks taking into account the reliability of D2D transmissions.
Using tools from stochastic geometry, we provide a closed-form approximation of cache-aided throughput, which measures the density of successfully served requests by local device caches, and we obtain the optimal caching probabilities with numerical optimization. Compared to the cache-hit-optimal case, the optimal caching probabilities obtained by cache-aided throughput optimization show notable gain in terms of the density of successfully served user requests, particularly in dense user environments.  
\end{abstract}

\vspace{-0.1cm}
\section{Introduction}
Proactive caching in wireless networks has attracted enormous attention as a means to reduce cellular data traffic load by bringing content closer to end users before being requested. Besides the conventional ``cache the most popular content everywhere" strategy, in random wireless networks with spatially distributed network nodes, one widely used caching strategy is the probabilistic content placement, referred to as \textit{geographic caching} \cite{optimalcaching} and \textit{independent random caching} \cite{d2dcaching} in the literature.
Recently, caching placement in random wireless networks with Poisson distributed caching helpers has been studied in different scenarios \cite{helper, caching_helper}. 
In contrast to the conventional optimization targeting in cache hit probability maximization \cite{d2dcdn}, in \cite{channel_diversity} the authors studied the optimization of the average success probability of content delivery in stochastic wireless caching helper networks in the noise-limited regime. The optimal content placement by maximizing the density of successful receptions was recently studied in \cite{maximum_dsr}. It is worth noticing that in D2D assisted wireless networks, caching at user devices leads to a special case of \textit{cache hit} when the requested content is cached within the device itself, which is often overlooked in the literature of wireless D2D caching networks. To the best of our knowledge, this special case was only addressed in \cite{d2d_optimalcaching}, referred to as \textit{self offloading}, where authors studied optimal probabilistic caching in a two-tier caching network by maximizing the offloading probability, which is equivalent to the cache hit probability in a general sense.

In this letter we investigate probabilistic caching placement in a stochastic wireless D2D caching network with two different objectives: 1) to maximize the cache hit probability, which is the probability that a random user request can be served locally, either by its own cache or by its neighbor devices within a certain distance, and 2) to maximize the cache-aided throughput, which is the average density of successfully served requests by local caches, including when a user request is served by its own device or by a nearby device through D2D transmission. The optimal solutions and their performances are evaluated, which shows that the reliability of D2D transmission plays a critical role in the optimal caching decisions in order to achieve the best throughput-related performance.

 \vspace{-0.1cm}
\section{Network Model} 
\label{section:network_model}
We consider a wireless D2D caching network where mobile user devices are modeled by a homogeneous Poisson Point Process (PPP) $\Phi_{\text{u}}$ with intensity $\lambda_{\text{u}}$. 
Each mobile user has probability $\rho\in[0,1]$ to be active, i.e., making an active request for a file, and the ``inactive" devices will serve as potential D2D transmitters.\footnote{This setting is based on the assumption that a user cannot serve nearby devices when it is waiting to be served.} Therefore, the distributions of receivers and potential transmitters follow homogeneous PPPs $\Phi_{\text{u}}^{{\text{r}}}$ and $\Phi_{\text{u}}^{{\text{t}}}$ with intensity $\rho\lambda_{\text{u}}$ and $(1-\rho)\lambda_{\text{u}}$, respectively. Each device has cache memory size $M_{\text{d}}$ and the files are assumed to have equal unit size.
Assuming a finite content category $\mathcal{F}=\{f_1, \cdots, f_N\}$, where $f_i$ is the $i$-th most popular file and $N>M_{\text{d}}$ is the library size. The content popularity follows the Zipf distribution, which is widely used in the literature \cite{optimalcaching, d2dcaching, helper, caching_helper, d2dcdn, channel_diversity, maximum_dsr, d2d_optimalcaching}, i.e., the request probability of file $f_i$ is 
\vspace{-0.05cm}
\begin{equation}
p_{i}=\frac{1}{i^\gamma  \sum\limits_{j=1}^{N}j^{-\gamma}},
\vspace{-0.05cm}
\end{equation}
where $\gamma$ is the shape parameter, which defines the skewness of the popularity distribution. 
We apply the \textit{geographic caching} strategy proposed in \cite{optimalcaching} on user devices, i.e., each user device independently caches file $f_i$ with a certain probability $q_{i}$. Denote $\mathbf{q}=[q_1,\ldots, q_N]$ the caching probabilities of file $i\in[1,N]$, we have $\sum\limits_{i=1}^{N}q_i\leq M_{\text{d}}$ due to the cache storage limit. As a result of independent thinning, the distribution of user devices who has the file $f_i$ follows a homogeneous PPP with intensity $q_i (1-\rho)\lambda_{\text{u}}$.

When an active user requests for a file in $\mathcal{F}$, a \textit{cache hit} event may happen in two cases: 
\begin{itemize}
	\item \textbf{Case 1: self-request}, when the requested file is cached in its own device.
	\item \textbf{Case 2: D2D cache hit}, when the requested file is not cached in its own device, but in one of its nearby devices within a certain distance $R_{\text{d}}$. If there is more than one D2D transmitters which has the requested file, the file is transmitted from the nearest one. 
\end{itemize}
In the case where the requested file can not be found in local device caches, the file is downloaded from the core network to the nearest base station through the backhaul, and then transmitted to the user. Here we assume D2D  \textit{overlaid} cellular networks, i.e., D2D and cellular communications thus, cross-tier interference does not exist.

\section{Performance Metrics and Analysis}
In this section we define the cache hit probability and the cache-aided throughput as the main performance metrics.
Note that the case of finding the requested file of a device in its own cache storage is often overlooked in helper-based D2D caching networks in the literature. This is the major difference between the cache-related performance study in this letter and prior work in \cite{caching_helper, d2dcdn,channel_diversity}.
\vspace{-0.3cm}
\subsection{Cache Hit Probability}
The cache hit probability is the probability of a random active user to find its requested file in local caches, including the \textit{self-request} and the \textit{D2D cache hit} cases. 

\subsubsection{Self-Request}
Denoting by $p_{\text{self}}^{\text{d}}$ the self-request probability of a random user is given by
\vspace{-0.2cm}
\begin{equation}
p_{\text{self}}=\sum\limits_{i=1}^{N} p_{i} q_{i}.
\label{self_request}
\vspace{-0.2cm}
\end{equation}
In this case, the request can be handled without the assistance of other devices.

\subsubsection{D2D Cache Hit}
As a result of the probabilistic caching, the probability to find a file cached inside a certain area strongly depends on the popularity order of the file and the area size. 
When a user requests for file $f_i$, the probability to find it cached in the devices within distance $R_{\text{d}}$ is \cite{d2d_vs_sc, d2d_optimalcaching}
\vspace{-0.15cm}
\begin{equation}
p_{\text{hit},i }^{\text{d}}=1-e^{-\pi (1-\rho)\lambda_{\text{u}}q_{i} R_{\text{d}}^{2}}.
\vspace{-0.15cm}
\end{equation}
Averaging over all the files in the content library $\mathcal{F}$, we have the D2D cache hit probability $p_{\text{hit}}^{\text{d}}=\sum\limits_{i=1}^{N} p_{i} \left(1-q_{i}\right)p_{\text{hit},i }^{\text{d}}$, thus,
\vspace{-0.2cm}
\begin{equation}
p_{\text{hit}}^{\text{d}}=\sum\limits_{i=1}^{N} p_{i} \left(1-q_{i}\right)\left(1-e^{-\pi (1-\rho)\lambda_{\text{u}}q_{i} R_{\text{d}}^{2}}\right).
\label{neighbor_service}
\vspace{-0.2cm}
\end{equation} 

The total cache hit probability is given by $p_{\text{hit}}=p_{\text{self}}+p_{\text{hit}}^{\text{d}}$, after replacing \eqref{self_request} and \eqref{neighbor_service}, we have 
\vspace{-0.2cm}
\begin{equation}
p_{\text{hit}}=1-\sum\limits_{i=1}^{N} p_{i} \left(1-q_{i}\right)e^{-\pi (1-\rho)\lambda_{\text{u}}q_{i} R_{\text{d}}^{2}}.
\label{cache_hit_proba}
\vspace{-0.2cm}
\end{equation}

\vspace{-0.5cm}
\subsection{Cache-Aided Throughput}
Unlike the conventional definition of network throughput, which measures the average number of information successfully transmitted over the network region, here we are interested in studying the average number of requests that can be successfully and simultaneously handled by the local caches per unit area, namely the cache-aided throughput (per area). 

Assume that the transmission of each file with equal size takes the same amount of time, one slot for instance. 
In the self-request case, the request is automatically served with probability one, while in the D2D cache hit case, the success probability of content delivery depends on the received signal-to-interference-plus-noise ratio (SINR). Thus, we have the cache-aided throughput given by 
\vspace{-0.15cm}
\begin{equation}
\label{throughput}
\mathcal{T}=\rho \lambda_{\text{u}} \left[\sum\limits_{i=1}^{N} p_i q_i\cdot 1+\sum\limits_{i=1}^{N} p_i (1-q_i) p_{\text{hit},i}^{\text{d}}\cdot p_{\text{suc},i}^{\text{d}} \right],
\vspace{-0.1cm}
\end{equation}
where $p_{\text{suc},i}^{\text{d}} $ is the success probability of D2D transmission for file $f_i$, $\rho\lambda_{\text{u}}$ is the density of user requests in a given time slot.
Without loss of generality, conditioning on having a typical active user $k$ to be served at the origin, the received SINR at the typical receiver is given by
\begin{equation*}
{\rm SINR}_k=\frac{P_{\text{d}}|h_{k,k}|^2 d_{k,k}^{-\alpha}}{\sigma^2+\sum_{j\in \Phi_{\text{t}}^{\text{d}}\backslash\left\{k\right\}} P_{\text{d}}|h_{j,k}|^2 d_{j,k}^{-\alpha}},
\end{equation*}
where $\Phi_{\text{t}}^{\text{d}}$ denotes the set of active D2D transmitters; $P_{\text{d}}$ denotes the device transmission power; $h_{j,k}$ denotes the small-scale channel fading from the transmitter $j$ to the receiver $k$, which follows $\mathcal{CN}(0,1)$ (Rayleigh fading); $d_{j,k}$ denotes the distance from the transmitter $j$ to the receiver $k$; $\sigma^2$ denotes the background thermal noise power. 

A file requested by a random user in $\Phi_{\text{r}}^{\text{u}}$ will be found within its nearby devices, but not in its own device with probability $p_{\text{hit}}^{\text{d}}$, as given in \eqref{neighbor_service}. Thus, the density of cache-assisted D2D transmissions is $\rho \lambda_{\text{u}} p_{\text{hit}}^{\text{d}}$.\footnote{Note that multiple users might find the same nearest D2D transmitter. In this case the transmitter will multicast the file to the receivers.} Although the distribution of the active D2D transmitters $\Phi_{\text{t}}^{\text{d}}$ is not a homogeneous PPP, the average density of $\Phi_{\text{t}}^{\text{d}}$ can still be approximated by
\vspace{-0.15cm}
\begin{equation}
\lambda_{\text{t}}^{\text{d}} \approx \rho \lambda_{\text{u}} p_{\text{hit}}^{\text{d}}.
\label{lambdatd}
\vspace{-0.15cm}
\end{equation}

When file $f_i$ is requested by the typical user, we denote $d_{i}$ the distance to the nearest device who has $f_i$ cached, and approximately consider $\Phi_{\text{t}}^{\text{d}}$ as a homogeneous PPP with intensity $\rho \lambda_{\text{u}} p_{\text{hit}}^{\text{d}}$. For a given SINR target  $\theta$ of successful D2D transmission, the D2D success probability is given as 
\vspace{-0.1cm}
\begin{align}
p_\text{\text{suc},i}^{\text{d}} &=\mathbb{P}\left[\frac{P_{\text{d}}|h_{k,k}|^2 d_{i}^{-\alpha}}{\sigma^2+\sum_{k\in \Phi_{\text{t}}^{\text{d}}\backslash\left\{k\right\}} P_{\text{d}}|h_{j,k}|^2 d_{j,k}^{-\alpha}} >\theta\right]     \nonumber\\
&=\mathbb{P}\left[|h_{k,k}|^2 >\frac{\theta d_{i}^{\alpha}}{P_{\text{d}}} \left(\sigma^2+\sum_{k\in \Phi_{\text{t}}^{\text{d}}\backslash\left\{k\right\}} P_{\text{d}}|h_{j,k}|^2 d_{j,k}^{-\alpha} \right)\right]   \nonumber \\
&\mathop{=}\limits^{(a)}\mathbb{E}_{d_{i}}\left[\mathcal{L}_{I_{\text{d}}}\left(\theta d_{i}^\alpha\right)  \cdot \exp\left(-\theta \sigma^2 d_{i}^\alpha/P_{\text{d}}\right) \right] \nonumber \\
&\mathop{=}\limits^{(b)}\mathbb{E}_{d_{i}}\left[\exp\!\left(\!-\frac{\pi \rho \lambda_{\text{u}} p_{\text{hit}}^{\text{d}} d_{i}^2 \theta^{\frac{2}{\alpha}}}{\sinc(2/\alpha)} \!\right) \exp\left(-\theta \sigma^2 d_{i}^\alpha/P_{\text{d}}\right) \right] \nonumber \\
&=\int_{0}^{\infty}\!\!\!f_{d_{i}}(r)\exp\!\left(\!-\frac{\pi \rho \lambda_{\text{u}} p_{\text{hit}}^{\text{d}} r^2 \theta^{\frac{2}{\alpha}}}{\sinc(2/\alpha)} \!\right)e^{-\frac{\theta \sigma^2 r^\alpha}{P_{\text{d}}}}\text{d}r,
\label{success proba}
\end{align}
where 
$\mathcal{L}_{I_d}(s)=\mathbb{E}\left[\exp\left(-s I_d\right)\right]$ is the Laplace transform of interference $I_d=\sum_{k\in \Phi_{\text{t}}^{\text{d}}\backslash\left\{k\right\}} |h_{j,k}|^2 d_{j,k}^{-\alpha} $, and $f_{d_{i}}(r)$ is the probability density function (PDF) of D2D distance $d_{i}$, when $f_i$ is requested by the typical user. 
Here, (a) follows from the complementary cumulative distribution function (CCDF) of $|h_{k,k}|^2$, which is exponentially distributed with unit mean value; (b) follows from the probability generating functional (PGFL) of PPP \cite{Haenggi}.

Conditioning on $d_{i}\leq R_{\text{d}}$ as a result of the maximum D2D distance, the PDF of $d_{i}$ is given by 
\begin{equation}
f_{d_{i}}(r)=\left\{
\begin{array}{rcl}
 &\frac{2\pi (1-\rho)\lambda_{\text{u}}q_i r}{1-e^{-\pi (1-\rho)\lambda_{\text{u}} q_{i} R_{\text{d}}^2}} e^{-\pi(1-\rho)\lambda_{\text{u}}q_i r^2}  & 0 \leq r \leq R_{\text{d}}\\
& 0   & r> R_{\text{d}}.
\end{array} \right.
\label{eq:fds}
\end{equation}

Substituting \eqref{success proba} and \eqref{eq:fds} in \eqref{throughput}, we obtain the cache-aided throughput averaged over all the files in the content library.

\vspace{-0.15cm}
\section{Optimization of Probabilistic Caching Placement}
In this section we study the optimal caching probabilities $\mathbf{q}=[q_1,\ldots, q_N]$ by cache hit maximization and by cache-aided throughput optimization, respectively.
\vspace{-0.3cm}
\subsection{Cache Hit Maximization}
\label{sec:hit optimization}
Based on \eqref{cache_hit_proba}, the optimization problem for maximizing the cache hit probability is defined as 
\vspace{-0.2cm}
\begin{eqnarray}
\underset{\mathbf{q}} {\max} ~&& p_{\text{hit}}=1-\sum\limits_{i=1}^{N} p_{i} \left(1-q_{i}\right)e^{-\pi (1-\rho)\lambda_{\text{u}}q_{i} R_{\text{d}}^{2}}\\
\mbox{s.t.} 
&&  0 \leq q_i\leq 1 \text{ for }  i=1,\ldots,N \nonumber \\
&& \sum\limits_{i=1}^{N}q_i \leq M_{\text{d}}. \nonumber
\vspace{-0.1cm}
\end{eqnarray}
The second order derivative of the objective function is strictly negative, thus $p_{\text{hit}}$ is a concave function of $q_i$ for $i=1,\ldots,N$.
Consider the following Lagrangian function
\vspace{-0.15cm}
\begin{equation}
\begin{split}
\mathcal{L}(\mathbf{q}, \mu)=-1+\sum\limits_{i=1}^{N} p_{i} \left(1-q_{i}\right)e^{-\pi (1-\rho)\lambda_{\text{u}}q_{i} R_{\text{d}}^{2}}&\\
+\mu\left(\sum\limits_{i=1}^{N}q_i - M_{\text{d}}\right),&
\end{split}
\end{equation}
where $\mu$ is the non-negative Lagrangian multiplier. We solve this optimization problem by applying the Karush-Kuhn-Tucker (KKT) conditions. 
From $\frac{\partial \mathcal{L}}{\partial q_i}=0$, we have 
\vspace{-0.1cm}
\begin{equation}
q_i(\mu)\!=\!-\frac{\mathcal{W}\left\{\!\frac{\mu}{p_i}\exp\left[1+\pi(1\!\!-\!\!\rho)\lambda_{\text{u}}R_{\text{d}}^2\right]\!\right\}}{\pi (1-\rho)\lambda_{\text{u}}R_{\text{d}}^2}+\frac{1}{\pi(1\!-\!\rho)\lambda_{\text{u}}R_{\text{d}}^2}+1,
\label{qi_mi}
\end{equation}
where $\mathcal{W}$ denotes the Lambert W function \cite{lambert}.
Combined with the condition $0 \leq q_i\leq 1$, let $[x]^{+}=\max \{x,0\}$, we have
\begin{equation}
q_i^{\star}=\min\left\{[q_i(\mu^{\star})]^{+},1\right\},
\end{equation}
where $\mu^{\star}$ can be obtained by the bisection search method under the other KKT condition $\sum\limits_{i=1}^{N}q_i^{\star} = M_{\text{d}}$.
\vspace{-0.3cm}

\subsection{Cache-aided Throughput Maximization}
\label{sec:throughput optimization}
Due to the complicated expression of $\mathcal{T}$, the optimal caching probabilities that maximize the cache-aided throughput are difficult to obtain, even with numerical methods. Consider the following approximation
\vspace{-0.2cm}
\begin{equation}
\mathbb{E}_{d_{i}}[\exp\left(-\eta d_{i}^\delta \right)]\approx \exp\left(-\eta \mathbb{E}[d_{i}^2]^{\delta/2}\right) ,
\vspace{-0.2cm}
\end{equation}
the success probability $p_{\text{suc},i}^{\text{d}}$ in \eqref{success proba} can be approximated by
\vspace{-0.2cm}
\begin{equation}
\hat{p}_{\text{suc},i}^{\text{d}}\approx\exp\!\!\left[-\frac{\pi\rho \lambda_{\text{u}} p_{\text{hit}}^{\text{d}} \mathbb{E}[d_{i}^2]\theta^{2/\alpha}}{\sinc(2/\alpha)}\right]\exp\!\!\left[-\frac{\theta\sigma^2 \mathbb{E}[d_{i}^2]^{\alpha/2}}{P_{\text{d}}}\right].
\vspace{-0.1cm}
\label{psuc_approxi}
\end{equation}
From the PDF of $d_{i}$ in \eqref{eq:fds}, we can obtain $\mathbb{E}[d_{i}^2]$ as follows.
\begin{align}
\mathbb{E}[d_{i}^2]&=\int_{0}^{R_{\text{d}}} r^2 \frac{2\pi (1-\rho)\lambda_{\text{u}}q_i r}{1-e^{-\pi (1-\rho)\lambda_{\text{u}} q_{i} R_{\text{d}}^2}} e^{-\pi(1-\rho)\lambda_{\text{u}}q_i r^2} \text{d}r  \nonumber\\
&=\frac{1}{\pi(1-\rho)\lambda_{\text{u}}q_i}-\frac{R^2_{\text{d}}}{e^{\pi (1-\rho)\lambda_{\text{u}} q_{i} R_{\text{d}}^2}-1}.
\end{align}
When $q_i\rightarrow0$, we obtain $\lim\limits_{q_i\rightarrow 0}\mathbb{E}[d_{i}^2]=R_{\text{d}}^2/2$ by applying L'H\^{o}pital's rule.

Then we have the approximated cache-aided throughput as
\vspace{-0.4cm}
\begin{align}
\hat{\mathcal{T}}=&\rho \lambda_{\text{u}} \left[\sum\limits_{i=1}^{N} p_i q_i+\sum\limits_{i=1}^{N} p_i (1-q_i) p_{\text{hit},i}^{\text{d}}\cdot \hat{p}_{\text{suc},i}^{\text{d}} \right],
\label{throughput_definition}
\vspace{-0.2cm}
\end{align}
where $\hat{p}_{\text{suc},i}^{\text{d}}$ is given in \eqref{psuc_approxi}. Our objective is to find
$\mathbf{q}^{\star}=\underset{\mathbf{q}} {\max}  ~ \hat{\mathcal{T}}$,
subject to $0 \leq q_i\leq 1$ and  $\sum\limits_{i=1}^{N}q_i \leq M_{\text{d}}$.

This problem is non-convex as it can be seen numerically. Providing an analytical solution to this problem is difficult. Therefore for the the cache-aided throughput maximization we solve it numerically with Simulated Annealing. 

\vspace{-0.3cm}
\section{Numerical and Simulation Results}
\label{section_4}
For numerical evaluation, we consider the user density between $\lambda_{\text{u}}=[10^{-4}, 10^{-3}]$/m$^{2}$. $\rho=50\%$ of the users will request for a random file in $\mathcal{F}$ according to the request probabilities $\mathbf{p}=[p_1,\ldots,p_N]$, which follows the Zipf distribution with parameter $\gamma=\{0.5, 1.2\}$. The rest $50\%$ of users act as potential D2D transmitters helping to serve the user requests locally.
The device cache capacity is $M_{\text{d} }=2$ files. The content library has size $N=20$ files.\footnote{Note that in reality the content library size is very large. Here we take $N=20$ files to avoid high complexity of the optimization problem. Similar choices can also be found in \cite{channel_diversity} and \cite{d2d_optimalcaching}.} The D2D searching distance is $R_{\text{d}}=75$ m. The device transmission power and the background noise power are $P_{\text{d}}=0.1$ mW and $\sigma^2=-110$ dB, respectively. The target SINR of successful D2D transmissions is $\theta=0$ dB. 

\begin{figure}
	\centering
	\includegraphics[width=0.9\linewidth]{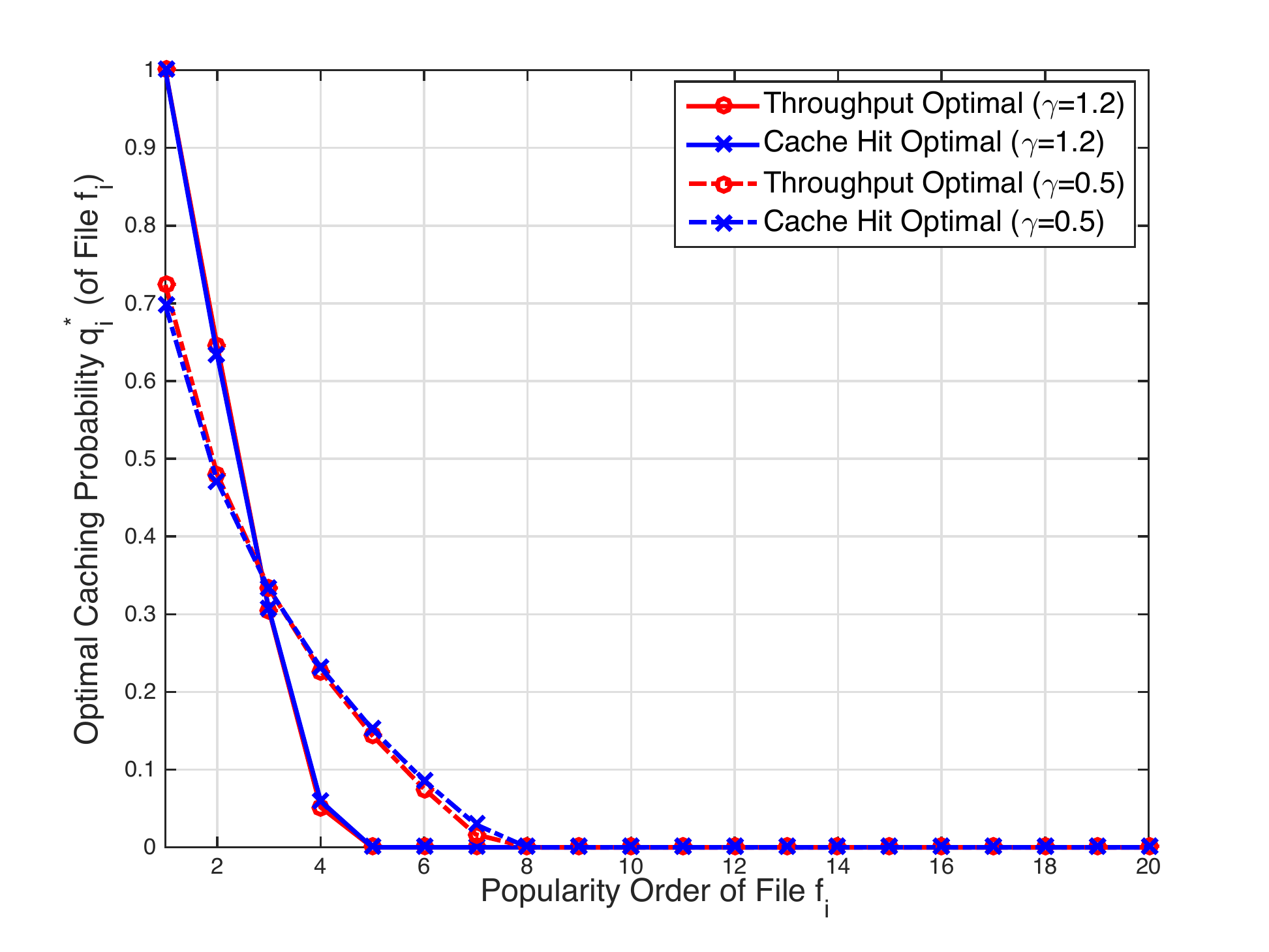}
		\vspace{-0.2cm}
	\caption{Optimal caching probabilities with sparse devices, $\lambda_{\text{u}}=10^{-4}$/m$^{2}$.}
	\label{fig:cacheprobasparse}
	\vspace{-0.5cm}
\end{figure}

\begin{figure}
	\centering
	\includegraphics[width=0.9\linewidth]{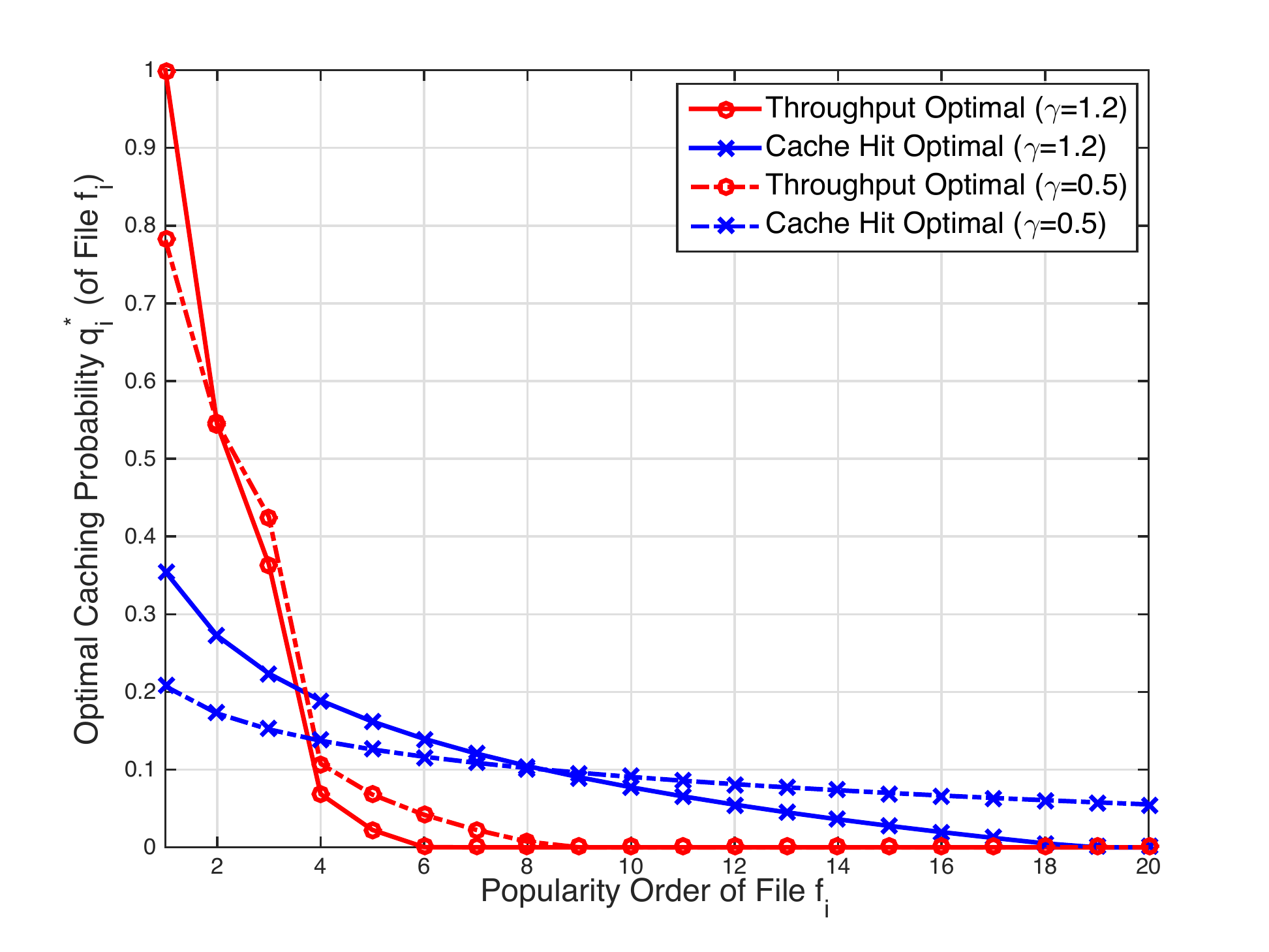}
		\vspace{-0.2cm}
	\caption{Optimal caching probabilities with dense devices, $\lambda_{\text{u}}=10^{-3}$/m$^{2}$.}
	\label{fig:cacheprobadense}
	\vspace{-0.5cm}
\end{figure}

In Fig.~\ref{fig:cacheprobasparse} and Fig.~\ref{fig:cacheprobadense} we compare the cache-hit-optimal (Section~\ref{sec:hit optimization}) and throughput-optimal (Section~\ref{sec:throughput optimization}) caching probabilities $\mathbf{q}^{\star}$ in sparse and dense user environments, respectively. The optimal caching probabilities of file $f_i$ for $i=1,\ldots,N$ are plotted as a function of the popularity order $i$. Interestingly, we observe that with sparse users, throughput-optimal and cache-hit-optimal caching probabilities are very close, while with dense users, each device tends to cache the most popular files with higher probability in order to increase the cache-aided throughput. For instance, in Fig.~\ref{fig:cacheprobadense}, $q_1^{\star}$, $q_2^{\star}$ and $q_3^{\star}$ in the throughput-optimal case are much higher than in the cache-hit-optimal case.
%, for both $\gamma=0.5$ and $\gamma=1.2$.
The intuition behind this is that \textit{in dense user regime, due to the excessive D2D interference that leads to very low D2D success probability, users' caching strategy tends to be more ``selfish" in the sense that self-request matters more than cache-assisted D2D transmission.}

\begin{figure}
\centering
\includegraphics[width=0.9\linewidth]{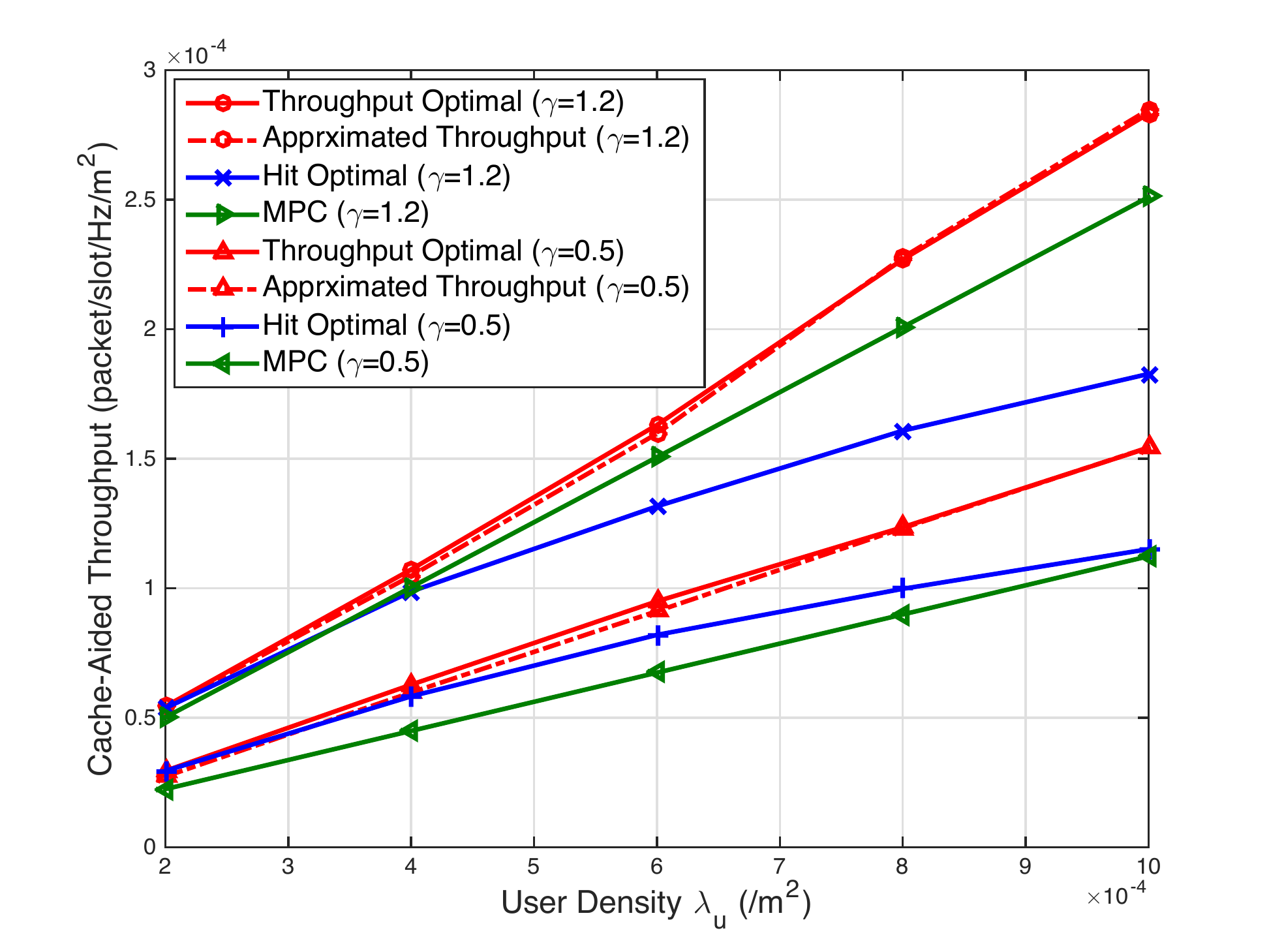}
	\vspace{-0.2cm}
\caption{Simulated cache-aided throughput vs. user device density $\lambda_{\text{u}}$. $\gamma=\{0.5, 1.2\}$}
\label{fig:thptcomparison}
	\vspace{-0.5cm}
\end{figure}

In Fig.~\ref{fig:thptcomparison} we plot the simulated cache-aided throughput obtained with the throughput-optimal caching probabilities. In order to validate the accuracy of the approximation we used in \eqref{psuc_approxi}, we plot the theoretical values of the approximated cache-aided throughput $\hat{\mathcal{T}}$, which turn out to have negligible error. 
For the comparison of different caching strategies, we also plot the simulated cache-aided throughput when applying $\mathbf{q^{\star}}$ obtained with cache hit probability optimization and with the conventional ``cache the most popular content" (MPC) strategy.
It is obvious that with the throughput-optimal strategy that is aware of the D2D success probability, the achieved cache-aided throughput can be significantly improved compared to the cache hit optimization and the MPC strategies. The gain is more profound in the dense user regime. Another interesting remark is that with dense users and highly concentrated content popularity ($\gamma=1.2$), the cache-aided throughput with MPC gives better performance than the cache-hit-optimal case, meaning that it is more beneficial to increase the chance of ``self-request" than increasing the total cache hit ratio. As summary, our results validate the necessity of taking into account the transmission reliability of cache-aided D2D communication while searching for the optimal content placement. 

\section{Conclusions}
\label{section_5}
In this letter, we studied probabilistic caching placement in stochastic wireless D2D caching networks with two objectives: maximizing cache hit probability and maximizing the cache-aided throughput that is defined by the density of successfully served user requests. The main takeaway message is that, in additional to the cache hit probability, the success probability of content delivery is also a critical factor that needs to be taken into account in the optimal caching placement in order to improve the throughput performance in wireless D2D caching networks.

\bibliographystyle{IEEEtran}
\bibliography{bib_ref}

\end{document}